\documentclass[letterpaper,12pt,final]{iopart}
\usepackage[utf8]{inputenc}
\usepackage{graphicx} 
\usepackage{subfigure}
\usepackage{wrapfig}
\usepackage{lineno}
\usepackage{multirow}

\usepackage{paralist}
\usepackage[colorlinks=true]{hyperref}
\hypersetup{pdftitle={}}

\begin{document}
\title{Introducing particle physics concepts through visual art}
\author{I.~Andrews$^{1}$, K.~Nikolopoulos$^{2}$}
\address{$^{1}$ In-Public, Community arts group, Birmingham, United Kingdom}
\address{$^{2}$ School of Physics and Astronomy, University of Birmingham, B15 2TT, United Kingdom}
\ead{k.nikolopoulos@bham.ac.uk}
\date{March 2018}

\begin{abstract}
The development of a workshop using the language, techniques, and
processes of visual art to introduce particle physics concepts is
described. Innovative delivery methods committed to the interaction
and collaboration of different specialist areas are utilised, which --
in curriculum terms -- encourages connections to be made between
separate subjects to the benefit of both. Beyond enhancing
understanding about the nature of the microcosm, this approach aims to
stimulate a ``creative curiosity'' about the world.
\end{abstract}

\submitto{\PED}

\section{Introduction}
The discovery of the electron by J.~J.~Thomson in 1897~\cite{electron}
inaugurated an era of discoveries and ever deeper understanding of the
inner workings of the microcosm.  This culminated, 115 years later, in
the Higgs boson discovery that completed the Standard Model of
particle physics. Presently, as in the time of Thomson, there are
several open questions that require an experimental breakthrough to be
answered.

With this intellectually stimulating state of affairs the ability of
visual arts to engage and express could ignite the curiosity of a
younger audience in particle physics, independently of any studies
they may pursue in the future.
This was confirmed by the authors during an artistic
collaboration culminating in the exhibition ``The Sketchbook and the
Collider''~\cite{residency}, where it became apparent that despite obvious
differences both specialisms are concerned with making the invisible,
visible. Scientific developments have seen the ``everyday'' dissolve
into sub-atomic interactions only accessible by examining traces left
in an enabling medium. A process mirrored by the artist expressing
thoughts, emotions and insights through marks made and materials
manipulated.

\section{Workshop development} 
A main motivation for the workshop is to creatively embed scientific
understanding, inspire interest in particle physcis, and stimulate a
``creative curiosity'' about the world. In parallel, it is important
that the work produced and methods involved are artistically valid
and that the workshops would prove beneficial for students studying
art, providing potential coursework with only limited additions to the
work produced during the day. Thus, connections between the
separate subjects are encouraged and support is provided for
innovative curriculum design, promoting 
inter-disciplinary collaborations.
The experience comprises of a range of methods and materials used in
current fine art education at first year degree level.
The design was further informed by the creative pedagogical
features~\cite{EXETER} established within the CREATIONS
project~\cite{CREATIONS}.

In terms of physics, the aspiration is for the students to understand
basic concepts regarding research in particle physics and the
structure of the microcosm. Specifically, that knowledge is acquired
through experiments, where particles interact in the detector
sensitive material and the information is used to construct images
that are interpreted as physics processes. For the structure of the
microcosm, students explore the particle content of the Standard
Model. They learn that matter particles make up the world around us,
they are categorised as quarks and leptons, and are organised in three
families. The interactions are mediated via particles of different
properties, e.g. the gluon, while the special role of the Higgs boson
is also explained. Finally, students are introduced to the idea that our
understanding is incomplete and ever evolving.

An initial range of drawing exercises, gradually developing from a
traditional base to increasingly experimental, is followed by
sculpture, photography and performance, as outlined in
Table~\ref{tab:overallplan}. As the workshop unfolds, the placing of
the outcomes onto a single A1-sized ``ideas'' sheet leads the pupils
to take greater ownership of the developing product, which is further
enhanced by the annotation of the outcome with scientific
explanations. Conveying concepts briefly in their own words offers
another learning opportunity. The ``ideas'' sheet then becomes the
focus of a series of group critiques during the workshop.

In the following, the details of workshop delivery are presented, also
as a model for art and science teachers. The pilot implementation
phase focused on students in Key Stage 4, aged 14 to 16, in the
British education system. Approximately 110 students, up to 16 per
workshop, from schools in the West Midlands participated. Girls
represented 60\% of the participants.

\begin{table}[h]
    \centering
    \begin{tabular}{p{2cm}p{10cm}c}
        Activity & Description & Duration\\
& & (min) \\\hline
discussion & standard model particles and interactions & 20\\
discussion & art-science collaborative examples & 10\\
drawing & charcoal and putty rubber exercise & 45 \\
critique & reviewing the ``ideas'' sheet & 15\\
drawing & pen and pencil exercise & 20 \\
drawing &  experimental mark making exercise & 15 \\\hline
\multicolumn{3}{c}{Break}\\ \hline
discussion & open questions: dark matter and dark energy& 15\\
sculpture & materials manipulation and light-box experiments& 30\\
performance & devise and film shadow screen performance & 30\\
critique & reviewing the completed ``ideas'' sheet and films & 15\\\hline
    \end{tabular}
    \caption{Physics and visual art workshop lesson plan.\label{tab:overallplan}}
\end{table}

\section{Introducing Particle Physics concepts}

At the beginning of the workshop, the stage is set by introducing
basic concepts of particle physics: the Standard Model particles and
interactions. Following initial exercises, the students are exposed to
major open questions, aiming to stimulate their imagination and
underline science as an on-going effort beyond a textbook list of
facts. During this discussion ``subatomic
plushies''~\cite{particlezoo}, a cloth model for each particle, are
used to provide a visual anchor to the discussion.

\subsection{The Standard Model}
A video of an operating cloud chamber~\cite{cloud} with Uranium
mineral at its centre gives the opportunity to explain the basic
operating principles of the detector, along with some historic facts,
and a connection to cloud formation in the atmosphere. The idea that
we learn things about nature through experiments, and by
``visualising'' the particles that are otherwise invisible, is
introduced.

Subsequently, a model of the atom, consisting of ``subatomic
plushies'', is provided. Students at KS4 are familiar with the idea
that matter consists of atoms, with a dense nucleus made of protons
and neutrons, and electrons around them. It is found that very few
students are aware of the term ``quark''. Students are invited to
``open'' the proton and the neutron to find that they consist of up-
and down-quarks, along with gluons keeping them together. The
electromagnetic force and its carrier, the photon, are explained in
terms of what keeps the electron orbiting around the nucleus. The weak
interactions are introduced, by connecting to the video and discussing
what leads to Uranium decaying. This opportunity is taken to discuss
also neutrinos, which are intimately connected to the weak
interaction.

Having introduced most of the first generation particles along with
the force carriers, the discussion expands to the three generations of
matter, essentially heavier replicas of the first generation. Students
are confronted with open questions: ``Why three generations? Are there
more?''. Finally, the Higgs boson and its special role in the Standard
Model is introduced.

\subsection{Beyond the Standard Model}

Following the drawing exercises, students are confronted with the idea
that the Standard Model particles account for less than 5\% of the
matter-energy content of the universe. Some of the evidence for the
existence of Dark Matter and Dark Energy are presented. It is pointed
out that intensive current research attempts to answer these questions, and that younger people, like
themselves, will contribute in the future. Usually, there are several questions
during this part of the workshop. Often students find it fascinating and intriguing that one
could know something exists despite not knowing much about its nature.

\section{Drawing}
Following the introduction to the world of particle physics and
armed with the imagery of cloud and bubble chamber images, showing the
trails of particles and the debris of particle interactions, 
students embark on visualisation drawing exercises. These have a
traditional starting point, and gradually become
experimental.

\subsection{Charcoal}
\begin{wrapfigure}{l}{0.30\linewidth}
\centering
\includegraphics[width=0.90\linewidth]{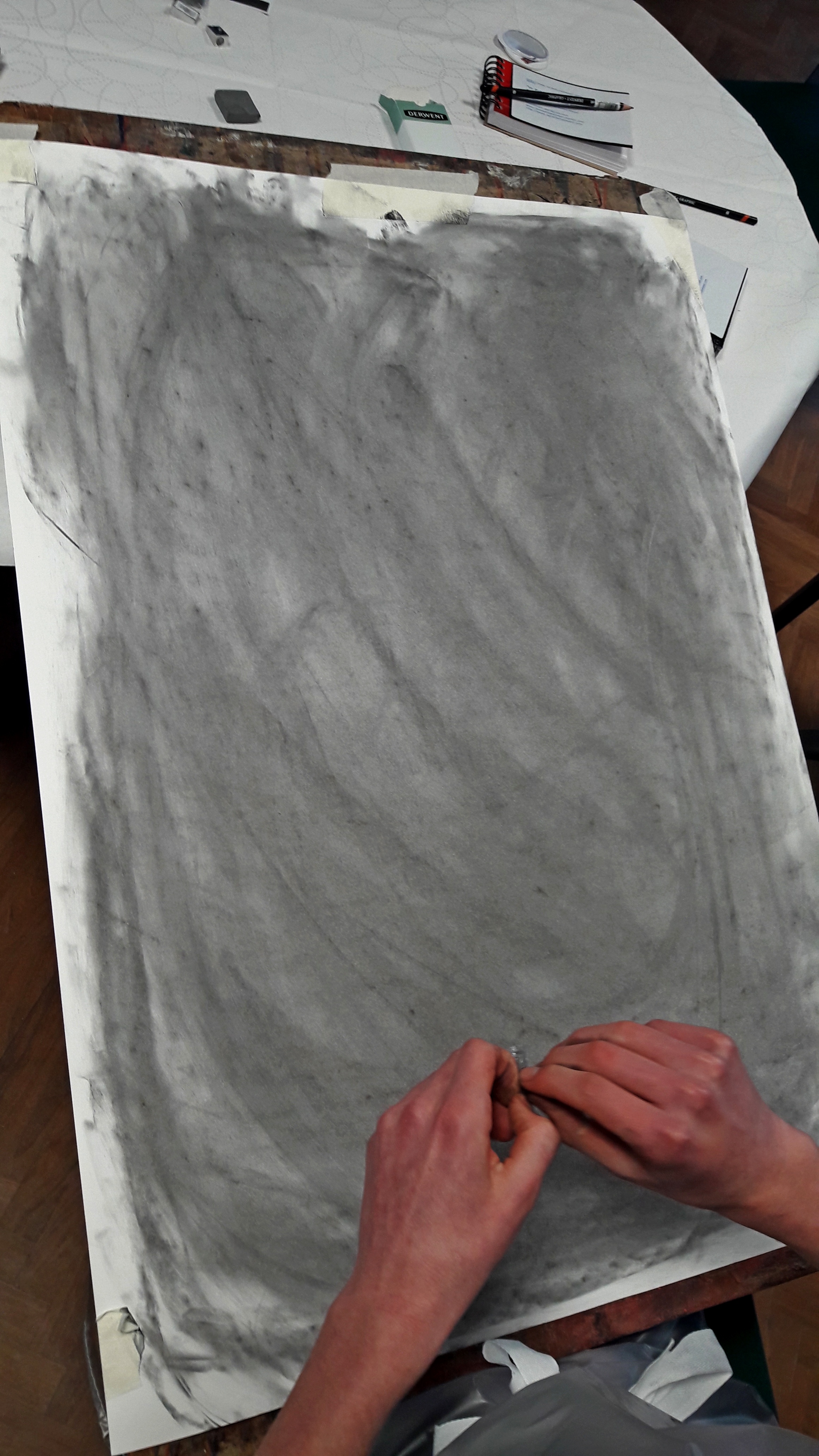}
\caption{Paper prepared with charcoal ground for first exercise with putty rubber.\label{fig:ground}}
\end{wrapfigure}
A charcoal ground is laid across the surface of a sheet of paper, as
shown in Fig.~\ref{fig:ground}. Putty rubbers are then used to draw
into the charcoal ground removing the charcoal and creating white
marks on the dark paper surface. In this context it represents the
particle interaction with the substance of the detector through which
the particle traces become visible. The idea of using interaction to
make visible the invisible is fundamental. 

This is a traditional technique, encouraging consideration of the
whole visual field when working on a composition. Students are
encouraged to consider overlapping the trails to enhance the
understanding that particles move and interact in all three spatial
dimensions, while the obtained images show a two-dimensional
projection of these events.

The laying down of the charcoal ground can also be used, depending on
the scientific level of the students involved, as a reference to the
Higgs field that is present everywhere in the universe, even in empty
space: the zero of white paper replaced with a surface of grey
charcoal.

This exercise is followed by a critique where students are
encouraged to discuss the artistic choices made in relation to the
scientific context. This is an opportunity to point out marks that may
closely resemble specific events as if they were actual bubble or
cloud chamber images~\cite{bubblechamber}:
\begin{inparaenum}[a)]
\item an incoming beam of particles;
\item ``vees'' from the decay of a neutral massive particle;
\item ``kinks'' when a charged particle decays; 
\item electron-positron pair production from a high energy photon; 
\item production of $\delta$-electrons, seen as trails perpendicular to a main track;
\item infering the direction of particles from the changing radius of repeated circles, as the particle losses energy; and
\item ``fiducial'' marks on the chamber walls, with accurately known positions, that are essential for event reconstruction.
\end{inparaenum}
Depictions that would lead to violation of fundamental physics laws,
e.g. conservation of the electric charge, are identified and
discussed. This enhances the visual equivalence between the artistic
outcome and particle physics experiment.

\subsection{Pen and Pencil}
The next set of exercises are more demanding artistically because they
relinquish traditional skills, move towards abstract thinking, and
require greater contextualisation.

\begin{figure}[h!]
\centering
\subfigure[]{\includegraphics[width=0.3\linewidth]{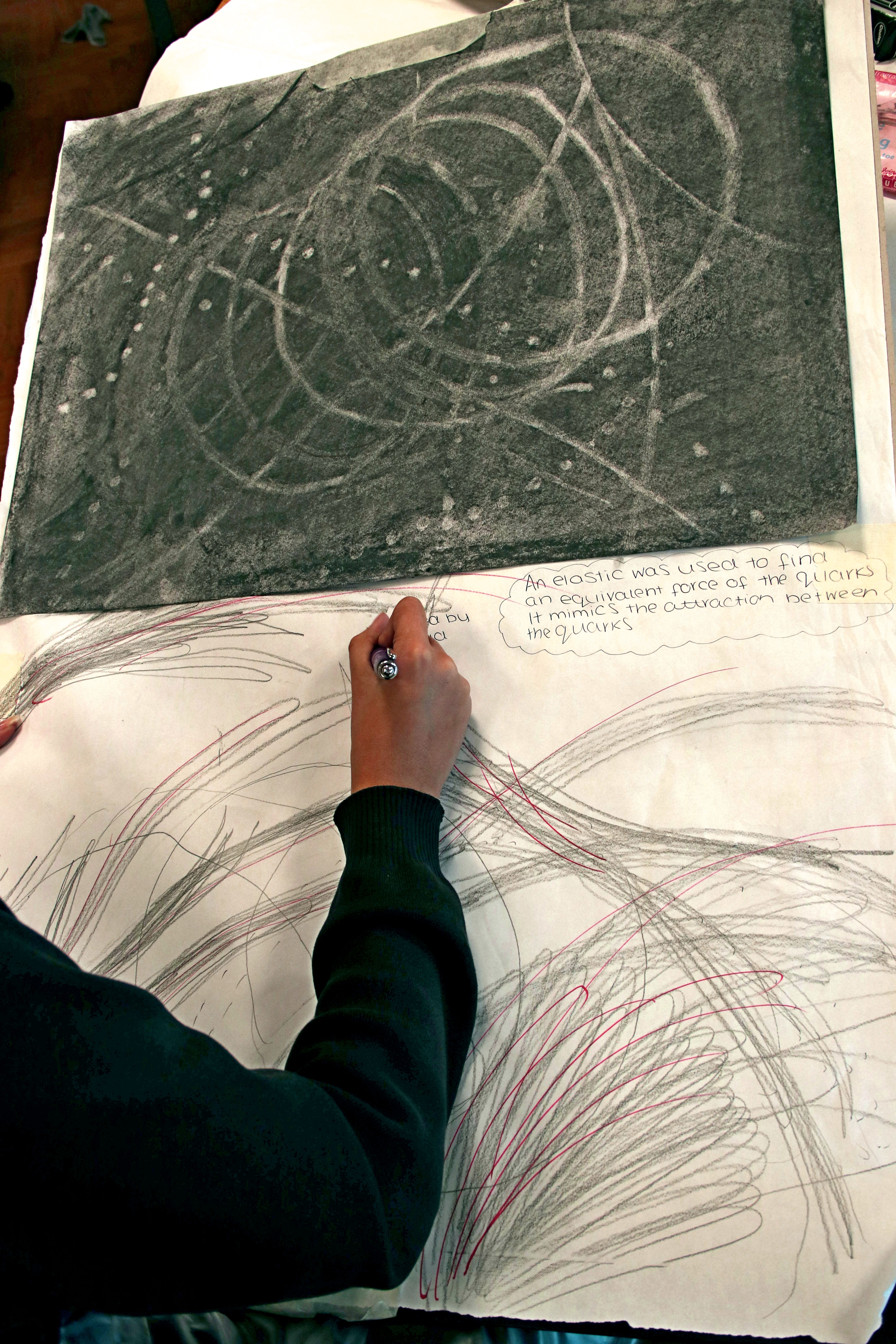}}
\subfigure[]{\includegraphics[width=0.3\linewidth]{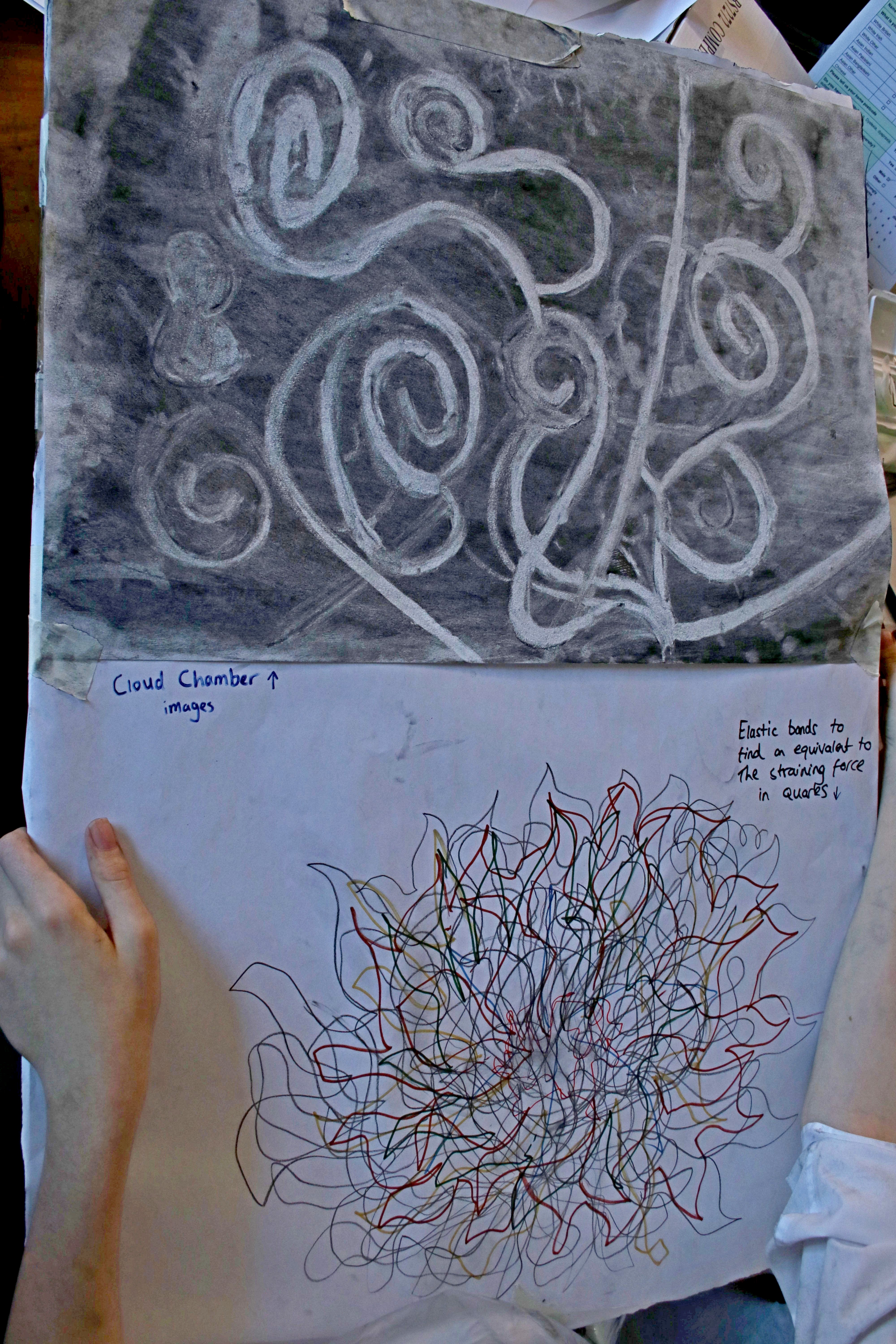}}
\caption{Examples of developing ideas sheet.\label{fig:ideassheet}}
\end{figure}

The increase in attractive force between two quarks as they are
separated is explored by drawing against the restrictive properties of
an elastic band wound around the pencil and other hand. This exercise
is expanded by students using different colours to further express
the increase in attraction and the use of more than one pen gripped in
the hand whilst drawing. Discussion takes place regarding 
``colour'' as the property, charge, that makes
particles interact through the strong interaction. Pencils, coloured
pencils, and felt tip pens are the most appropriate media in this
case. Examples of the work produced are shown in Fig.~\ref{fig:ideassheet}.

A collaborative approach to drawing is employed in this next exercise,
exploring the idea of particles interacting, decaying, and
transforming. Students create a drawing in a short 
time, 30 seconds or less, and then pass it to a neighbour who must
respond in an equally short time period. The stipulation is that each
participant must begin their drawing by working into the marks already
on the page. The drawing is continuously passed around the group until
all have added a transformative series of marks to the overall
composition. Media used in experimentation so far have been pencils,
coloured pencils, and felt tip pens as the speed of the exercise makes
any messier materials counter-productive. The time period depends on
the size of the group but the activity is designed to be fast to
reinforce the idea of the short lifespan of some of the particles
involved. This exercise is inspired from
the ``exquisite corpse'' collaborative drawings of the Surrealists.

The students can dislike the idea of another person intervening on
their art work. In a scientific context this is used to explain that particle physicists work
collaboratively, nowadays in very large teams from different countries, and that collaboration, and the ability to
communicate and exchange ideas is crucial for progress.

\subsection{Experimental mark making methods}

\begin{wrapfigure}{r}{0.4\linewidth}
\centering
\includegraphics[width=0.99\linewidth]{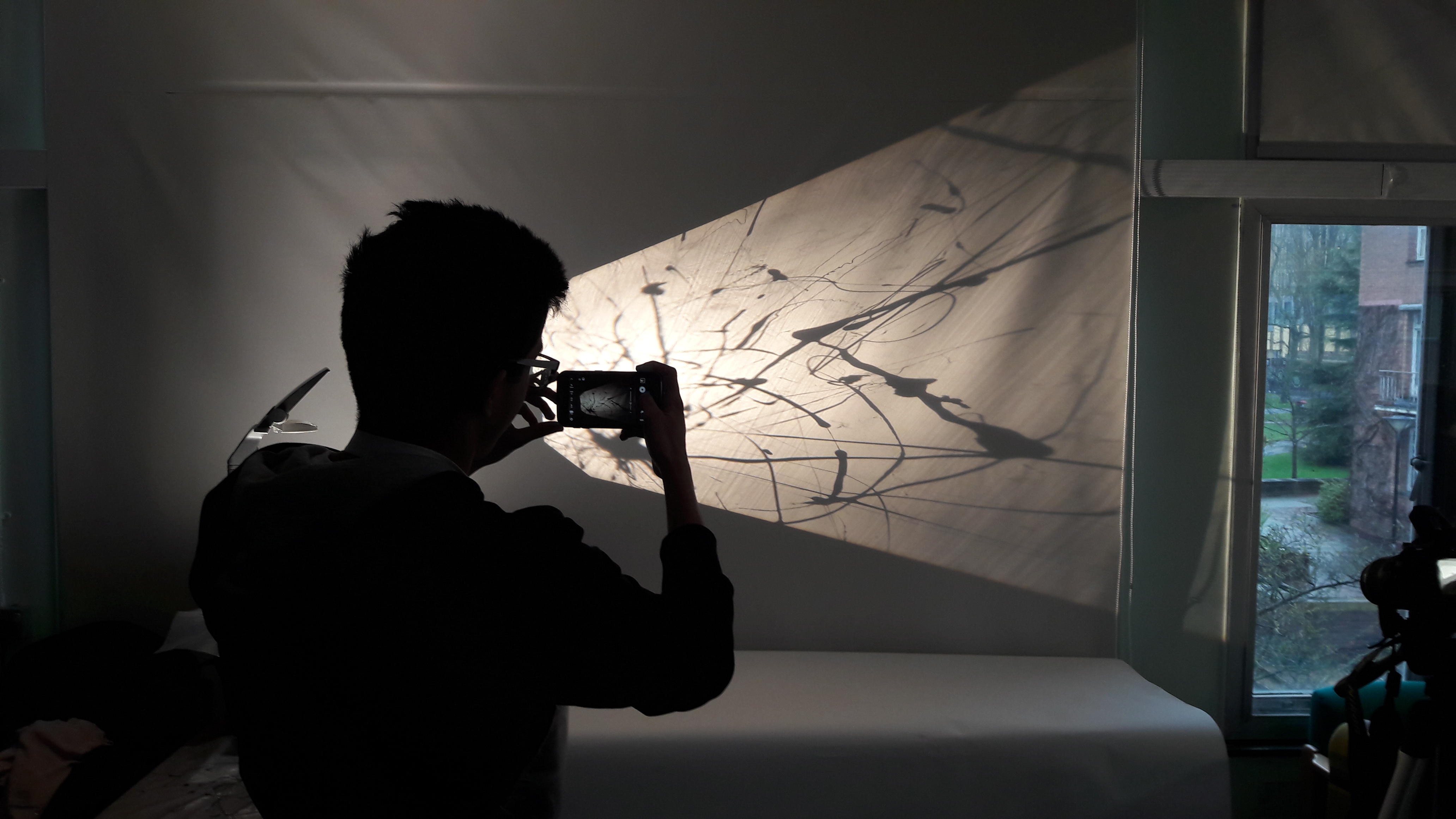}
\caption{Example of experimental mark making on transparent surface and projection.\label{fig:gluegun}}
\end{wrapfigure}
Continuing into the workshop, the use of glue guns as a drawing tool
takes the definition of drawing to an extreme that mirrors the
experimental nature of much of current visual art practice. It also
enables the creation of work on a range of surfaces which if
transparent, like polythene or plastic sheet, can be placed upon a
light-box and projected. The transformation of a small drawn starting
point that becomes a large light piece, that can be further changed by
distortions caused by the surface on which it is projected, as shown
in Fig.~\ref{fig:gluegun}, is used for the development of ideas
regarding the transformation and interaction of particles.

\section{Sculpture}
In this section pupils explore the sculpture techniques with
simple materials in three dimensions to re-emphasise some of the
particle physics concepts introduced and potentially explore new
ones. Materials have been restricted to various forms of wire,
string, tape, cocktail sticks and air-drying clay.

\begin{figure}[h!]
\centering
\subfigure[\label{fig:sculpturea}]{\includegraphics[width=0.50\linewidth]{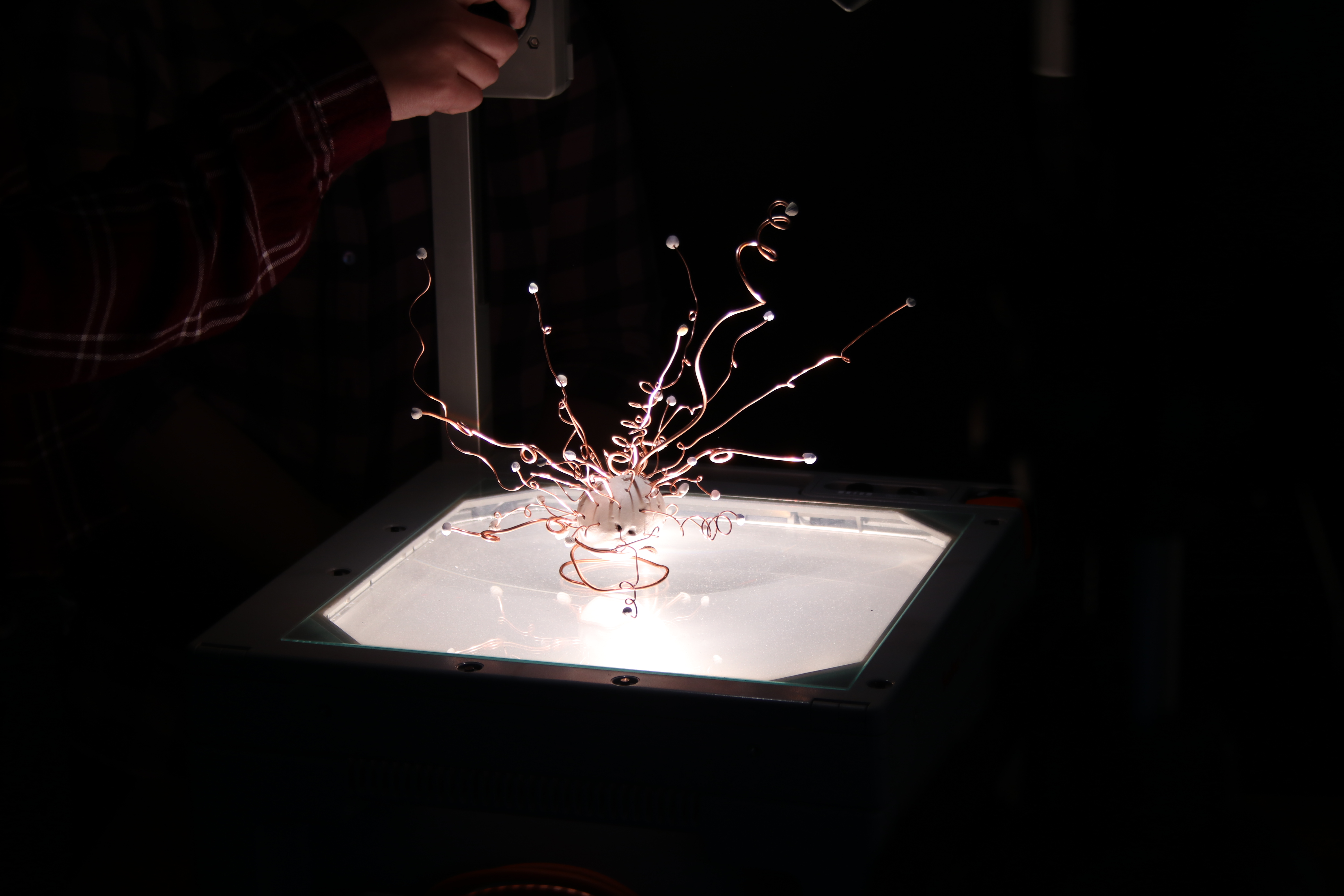}}
\subfigure[\label{fig:sculptureb}]{\includegraphics[width=0.335\linewidth, angle=90]{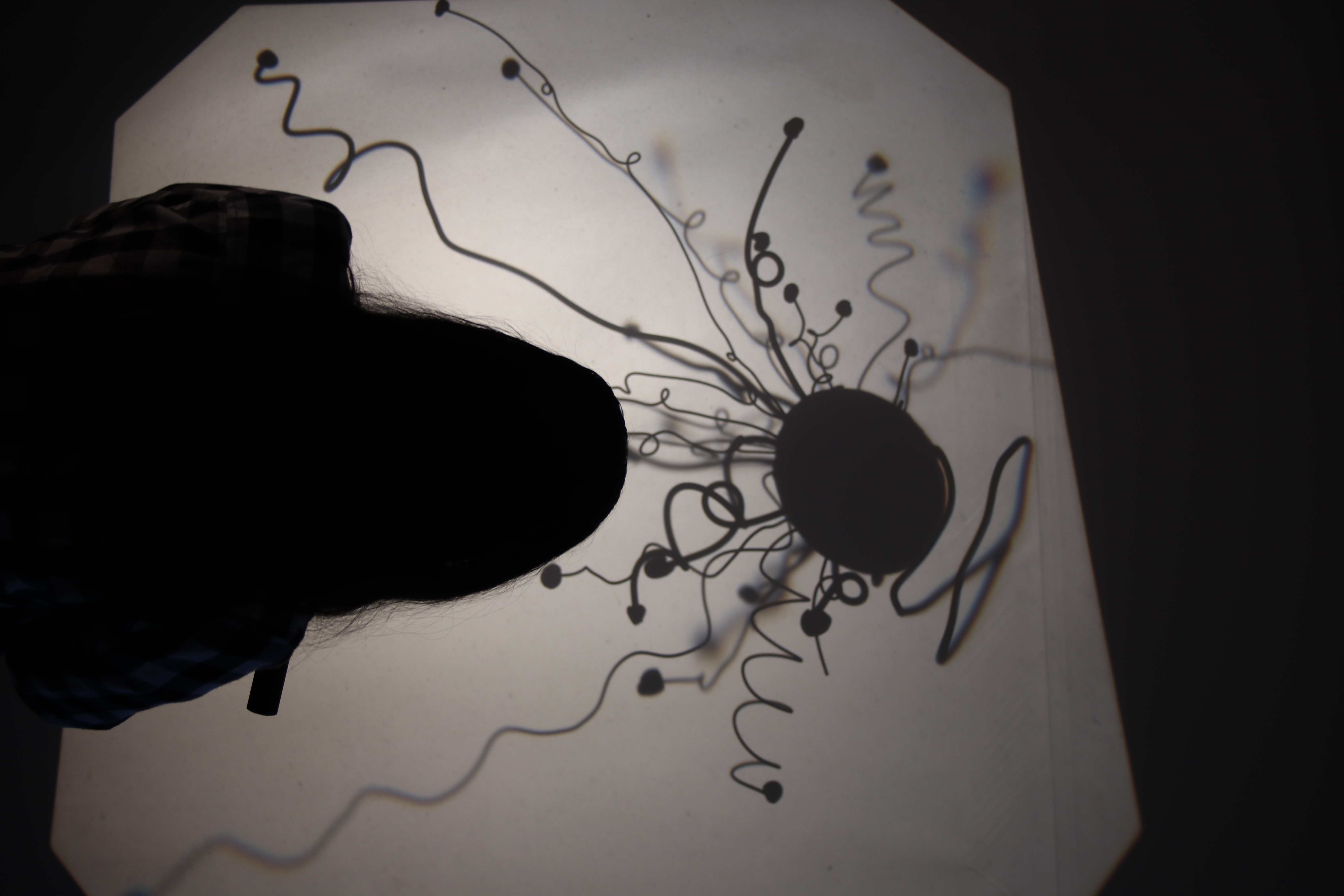}}
\caption{Example of \subref{fig:sculpturea} sculptural exercise and \subref{fig:sculptureb} resultant projection. Photograph by D.~Spathara.\label{fig:sculpture}}
\end{figure}

The drawn starting points transform from two dimensions into three
dimensions, with various forms of wire taking the place of drawn lines
and air-drying clay enabling drawn shapes to be translated into
form. The transformation of particles is discussed and when the
sculptural form is placed on the light-box and taken back into two
dimensions via projected light and shadow, the nature of
transformation and elements changing state can be introduced or
reinforced. The making of sculptures from some of the earlier drawings
re-iterated that particle traces are being created in a
three-dimensional space. An example of a sculpture is shown in
Fig.~\ref{fig:sculpturea} and its projection in two dimensions in
Fig.~\ref{fig:sculptureb}.

\section{Shadow screen performance}

Ideas that it would be challenging to include in a drawing can be
explored in a performance, by using the movement and interaction to
express and discuss activity in the particle world.  The students,
divided in two smaller groups, devise, perform, and film a short
performance piece using the movement and interaction of their own
bodies. Following discussion, the students allocate roles: direction,
sound, performance, camera etc. To avoid self-consciousness the
performance is behind a paper screen with an overhead projector
casting their shadows onto the screen.

Students consider the different particle physics
concepts to which they have been introduced. New concepts can be introduced at
this point to particularly able groups or older participants.
The concept is expressed through choreography with the students deciding on
the props and sounds that would reinforce the content, and
whether a background image can be produced, projected and moved on
the overhead projector in addition to the movement of the participants
themselves.

\begin{figure}
\centering
\includegraphics[width=0.45\linewidth]{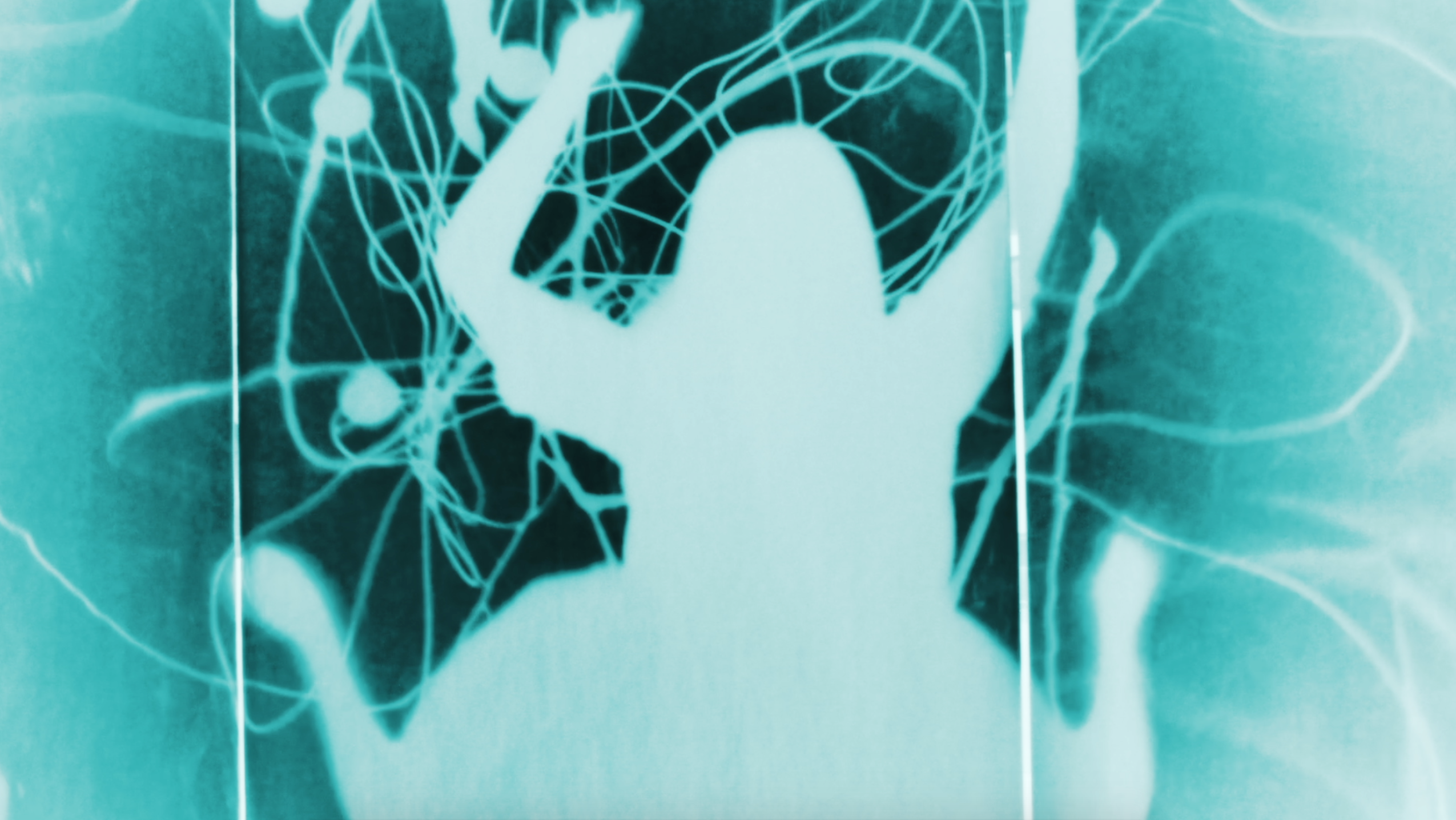}
\caption{Example of a still image from performance, following preliminary editing with students.\label{fig:film}}
\end{figure}

As an example, a performance designed to show the increasing
attraction when quark pairs are separated, involved two students pulling
on a binding wrapped around them in an increasingly agitated fashion
as the distance between them grew. The concept was reinforced by the
background image, produced during the glue gun experimentation, also
moving in an increasingly agitated fashion out of sync with the
movement of the performers. Combined with a soundtrack of rustling
paper becoming increasingly loud and frantic. The combination produced
a memorable sequence that made an impression to the students involved.

Depending on the abilities of the group it is often possible
to edit the footage produced to obtain a more resolved
film. Examples of simple interventions that can be applied are:
\begin{inparaenum}[a)]
\item reversing the tones, with light areas becoming dark and vice-versa, enabling the reinforcement of the idea of antimatter if relevant;
\item heightening the colour, making the footage very hot, can reinforce the idea of energy concentrating and dissipating; and 
\item changing the colour entirely, reinforceing the idea of the transformation of particles.
\end{inparaenum}
An example still image is shown in Fig.~\ref{fig:film}.

\section{Critique}
\begin{wrapfigure}{l}{0.50\linewidth}
\centering
\includegraphics[width=0.95\linewidth]{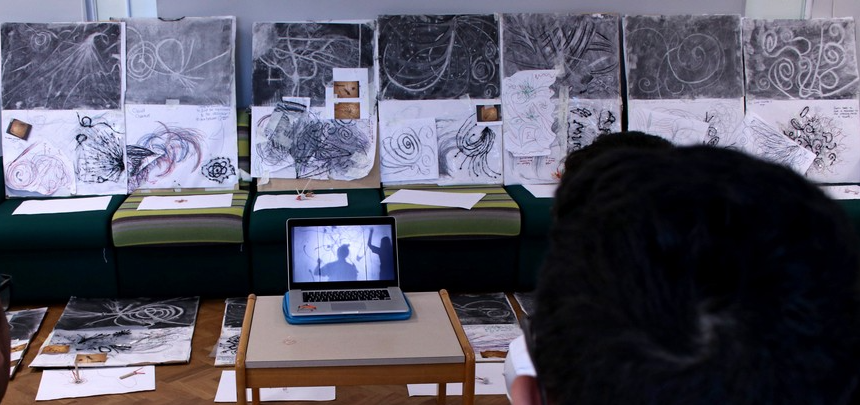}
\caption{The ``ideas'' sheets, sculptures, and filmes exhibited for discussion at group critique. Photograph by D.~Spathara.
\label{fig:critique}}
\end{wrapfigure}

The workshops conclude with a group critique of the
``ideas'' sheet. An example of the drawing exercises, the photographic evidence of the light
projections, the sculptures, and a viewing of the films produced, is presented in Fig.~\ref{fig:critique}.

This final discussion session is used to further reinforce
key concepts and to check for and dispel any misconceptions that may
have developed. This activity is also linked to the scientific
seminar and the idea that progress relies on scientists
presenting their results, discussing them, and potentially changing or improving
them following suggestions.

The full ``art school'' experience using the range of activities enables
cross connections to be made and scientific concepts to be reinforced
from one activity to another. For example the attraction between
quarks may be remembered by the use of the elastic band exercise but
also as the subject of a shadow screen performance.

\section{Evaluation and Feedback}

\begin{figure}[h!]
\centering
\includegraphics[width=0.60\linewidth]{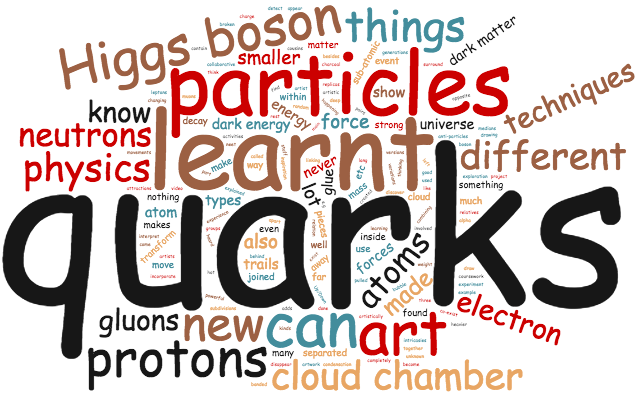}
\caption{Word cloud of the students' response to the question: ``Describe one thing that you learnt today''.
\label{fig:wordcloud}}
\end{figure}

During the workshop learning was evaluated informally through
one-to-one discussions. Students have shown growing understanding
regarding detector operation, the families of particles, and the
properties of the force carriers. They also discussed the concept that
had the largest impact on them, as shown in Fig.~\ref{fig:wordcloud},
with the term quark appearing prominently.

At the end of each workshop students were invited to comment on areas
of particular success and areas for improvement. At the same time
feedback was requested from the teachers accompanying the students.
Discussing physics through visual arts, led to an increased interest
in physics for four out of five students. Approximately 35\% of the
students found it more likely to study physics at university after
this event, while this fraction was 20\% among students that were not
originally inclined to study physics. Finally, after the workshop
three quarters of the students considered physics to be a more
accessible subject, including 65\% of students that initially did not
consider physics an accessible subject.

\section{Discussion and Future Directions}
The experience and feedback acquired during the pilot implementation phase is being reflected
upon to improve the workshop and plan the roll-out of a streamlined
version aiming to reach a larger audience. This could be achieved by
implementation at schools through collaboration of science and art
teachers, potentially through teacher training sessions.

The materials used are readily available, with the crucial point being
the imagination with which the materials are used, while the use
of more complex, expensive, and difficult to obtain materials is
likely to reduce the effectiveness of the workshop, as the cost
becomes a prohibiting factor. A possible exception is the film editing
which was done by the artist, and in the future by the art teacher,
involved, with the students advising on what filters and effects they
think are most appropriate. All of these activities require the
participants to use the clues given by the exercise, e.g. the
restrictive nature of the elastic band, and to make marks suggested by
the forces involved.

The workshop includes several strands of visual arts to maximise the
probability that each student will find inspiration. However, each of
these activities, appropriately expanded, could form a separate
workshop. This preserves the art school experience and has the added
benefits of increasing the involvement of each student with a given
art form, deploying more advanced techniques, maximising learning
time, and allowing for the creation of more complete artworks.  For
example the sculpture activitity could be expanded by introducing
additional materials and using larger areas, while keeping some links
to drawing through a preceding activity, as initial exploration and
design thinking prior to making. The shadow screen performance
activity could be extended by involving design drawing and
storyboarding prior to the production of the performance
itself. Provided sufficient resources are available, the editing could
become a singificant feature, done by the students themselves.

Although the workshop was initially focused on KS4 students, with
minimal changes it can be addressed to older or younger students, and
to larger class sizes without loss of its distinctive features. For
younger participants, for example, developing the sculptural unpacking
into a form of ``pass the parcel'' could be used as an interactive,
fun element whilst emphasising the unwrapping or splitting of the
particles down to their elementary constituents.  At the same time a
sitting configuration of participants, rather then the usual circular,
could be established in advance representing a specific particle
interaction, while, depending on age, Feynman diagrams could also
provide teaching aids and introduce specific scientific
visualisations.

\section{Summary}
An ``art school in a day'' workshop has been developed and
implemented, aspiring to introduce particle physics concepts to
students through the use of visual art, while stimulating a ``creative
curiosity'' towards the world. The workshop consists of drawing,
sculpture, and shadow screen performance. The students are introduced
to particle physics concepts new to them, and the art outcomes are discussed
both in relation to the scientific concepts and aesthetics. Building
on the promising evaluation of the pilot phase, future steps involve
focused implementation and expansion to different ages to allow for
larger classrooms, and the training of science and art teachers so
that they can deliver the workshop in their local schools.

\ack The authors would like to thank Sarah Fortes-Mayer, co-founder of
In-Public, and Dr Paul Thompson for their participation in the
implementation phase of the workshop. We thank Dr Maria Pavlidou for
her support as the University of Birmingham School of Physics and
Astronomy Liaison Officer and Dimitra Spathara for documenting
photographically the workshops. Feedback and suggestions of Dr Maria
Pavlidou and Dr Elizabeth Cunningham on the manuscript are gratefully
acknowledged. This work has been supported by the European Commission
as part of the ``CREATIONS - Developing an engaging science class
room'' support and coordination action (CREATIONS-665917).

\section*{References}
\bibliographystyle{ieeetr}
\bibliography{references}
\end{document}